\documentclass[twocolumn,superscriptaddress,floatfix,preprintnumbers,prl ,nofootinbib]{revtex4}
\usepackage[colorlinks=true,breaklinks=true]{hyperref}
\usepackage[utf8]{inputenc}
\hypersetup{allcolors=[rgb]{0.0 0.0 0.6},linkcolor=[rgb]{0.75 0.05 0.05}}
\usepackage{mathtools}
\usepackage[dvipsnames]{xcolor}
\usepackage{float}
\usepackage{siunitx}
\usepackage{letltxmacro}
\LetLtxMacro{\oldcite}{\cite}
\renewcommand{\cite}[1]{\mbox{\oldcite{#1}}}

\DeclareSIUnit\electronvolt{e\kern-.05em V}
\DeclareSIUnit\parsec{\text{pc}}
\DeclareSIUnit\clight{\text{\ensuremath{c}}}

\usepackage{orcidlink}

\long\def\exclude#1{}

\def\lya{Lyman-$\alpha$\,}

\def\fiso{f_{\rm iso}}

\newcommand{\beq}{\begin{equation}}
\newcommand{\eeq}{\end{equation}}

\def\ga{\,\,\raise0.14em\hbox{$>$}\kern-0.76em\lower0.28em\hbox
{$\sim$}\,\,}

\newcommand{\taureio}{\tau_{\mathrm{e}}}

\allowdisplaybreaks

\bibpunct{[}{]}{,}{n}{}{,}

\usepackage[colorinlistoftodos]{todonotes}

\begin{document}

\title{Post-inflationary axion constraints from the Lyman-$\alpha$ forest}

\author{Olga Garcia-Gallego\,\orcidlink{0009-0009-9003-7889}}
\email{og313@cam.ac.uk}
\affiliation{Institute of Astronomy, University of Cambridge, Madingley Road, Cambridge CB3 0HA, UK}
\affiliation{KICC - Kavli Institute for Cosmology Cambridge, Madingley Road, CB3 0HA Cambridge, United Kingdom}
\author{Vid Ir\v{s}i\v{c}\,\orcidlink{0000-0002-5445-461X}} \email{v.irsic@herts.ac.uk}
\affiliation{KICC - Kavli Institute for Cosmology Cambridge, Madingley Road, CB3 0HA Cambridge, United Kingdom}
\affiliation{Center for Astrophysics Research, Department of Physics, Astronomy and Mathematics, University of Hertfordshire, College Lane, Hatfield AL10 9AB, UK}
\author{Matteo Viel\,\orcidlink{0000-0002-2642-5707}}
\email{viel@sissa.it}
\affiliation{SISSA - International School for Advanced Studies, Via Bonomea 265, I-34136 Trieste, Italy}
\affiliation{IFPU, Institute for Fundamental Physics of the Universe, Via Beirut 2, I-34151 Trieste, Italy}
\affiliation{INFN, Sezione di Trieste, Via Valerio 2, I-34127 Trieste, Italy}
\affiliation{INAF - Osservatorio Astronomico di Trieste, Via G. B. Tiepolo 11, I-34143 Trieste, Italy}

\author{Martin G. Haehnelt\,\orcidlink{0000-0001-8443-2393}}
\affiliation{Institute of Astronomy, University of Cambridge, Madingley Road, Cambridge CB3 0HA, UK}
\affiliation{KICC - Kavli Institute for Cosmology Cambridge, Madingley Road, CB3 0HA Cambridge, United Kingdom}

\author{James S. Bolton\,\orcidlink{0000-0003-2764-8248}}
\affiliation{School of Physics and Astronomy, University of Nottingham, University Park, Nottingham, NG7 2RD, UK}


\begin{abstract}
Among the most compelling cold dark matter candidates, the axion has recently been subject to a wide range of astrophysical studies aiming to constraints its properties. We present updated bounds on the isocurvature fraction, $\fiso$, which parameterizes the contribution of isocurvature perturbations induced by post-inflationary produced axion-like particles (ALPs) to the ordinary power spectrum. We use new simulations based on the Sherwood-Relics suite to fit high-resolution Lyman-$\alpha$ forest flux power spectrum data.  With the published noise model of the Lyman-$\alpha$ forest data, we find a tentative detection of $\fiso$ = ${0.0064^{+0.0012}_{-0.0014}}$ (68\% C.L), after accounting for the degenerate effect of IGM thermal evolution. With a more conservative modelling of the residual noise in the data, the upper bound is weakened to $\fiso< 0.0084$ (95\% C.L), which translates into an ALP temperature-independent mass $m_a > 1.73 \times 10^{-18}$eV. Our constraints are stronger than bounds derived from large-scale structure probes at higher and lower redshifts and are competitive with those derived from UV luminosity function data. Interestingly, the best current Lyman-$\alpha$ forest data prefers a non-zero contribution from isocurvature modes.
\end{abstract}

\maketitle

\textit{Introduction.}--- The existence of non-baryonic cold dark matter (CDM) explains the process of hierarchical structure formation and growth in our Universe. Cosmological observations have led to two preferred CDM candidates: thermally produced Weakly Interacting Massive Particles (WIMPs), with ongoing experimental searches targeting the TeV scale, and axions, a wave-like candidate emerging in Standard Model extensions and initially proposed to solve the strong CP problem of QCD \cite{quinn1977, weinberg1978, wilczek1978}. More generally, axion-like particles (ALPs) are well-motivated in string theory, where axions appear upon compactification as pseudo-Nambu-Goldstone bosons, with a broader range of masses and coupling constants, as well as in  grand unified theories (GUTs) \cite{chadha22, marsh15}. This, together with the unsuccessful detection of the WIMP, has led to ALPs gaining increasing attention in the recent years. 

The most commonly studied way to produce ALPs is the \textit{misalignment} mechanism \cite{preskill83, dine83, abott83}. In the early Universe, the axion field is offset from the minimum of its potential or vacuum value. When the axion's mass field, $m_{a}$, becomes comparable to the Hubble expansion rate,  $H$, the field rolls down to the minimum of the potential and starts oscillating coherently. The energy density of the axion field then mimics the evolution of CDM \cite{preskill83, dine83, turner83}. Whether the spontaneous global symmetry breaking that establishes the axion in the first place occurs before or after inflation, will have important cosmological and astrophysical implications \cite{feix20}. 

In the post-inflationary case, since the initial vacuum misalignment angle fluctuates across casually disconnected patches, the ALPs generate a spectrum of white-noise or isocurvature perturbations, which were first discussed by \cite{peebles1970, efstathiou1980} and constrained by \cite{enquivist2000} from the CMB anisotropies. Observations from Planck later demonstrated that these isocurvature modes only contribute subdominantly to the CMB power spectrum (e.g. \cite{beltran2004, dunkley2005}). The CMB can also be combined with other large-scale structure (LSS) probes at high enough redshifts, such that the matter power spectrum is not significantly affected by the growth of non-linear perturbations. Since the contribution of isocurvature perturbations is constant power at small scales, their impact on the total power spectrum is characterized by $\fiso$, the ratio of isocurvature to adiabatic perturbations at $k_{\star}=0.05$ Mpc$^{-1}$. This parameter has been constrained to  $\fiso <$ 0.31 (2$\sigma$) with Planck data \cite{feix19}.
The bound can be translated to $m_a$ limits ($m_a > 10^{-20}-10^{-16}$ eV), where the four-order-of-magnitude difference is due to details of the mass-temperature dependence. Compared to the QCD axion, the range of ALPs masses becomes more uncertain since $m_a$ is no longer generated by QCD effects thereby becoming an independent parameter of the model. A tighter bound of $m_a \geq 10^{-19}$ eV was found in subsequent work with CMB+BAO+SZ cluster counts \cite{feix20}. 
The other relevant work for these ALP models is that from \cite{irsic19}, who discussed the constraining power from different early structure formation probes, including Lyman-$\alpha$ forest observations. The latter, obtained by translating primordial black holes (PBHs) bounds from \cite{murgia19} ($\fiso <$ 0.004 (2$\sigma$)) into ALP bounds, $m_a > 2 \times 10^{-17}$ eV, was found the most sensitive probe of all the observables considered.

Also relevant to this work, \cite{pavivek25} recently found a tentative detection of small-scale power enhancement when fitting the Lyman-$\alpha$ forest data from \cite{boera19} to simulations that included the effect of primordial magnetic fields (PMFs). The extra power signature also arises in models where PBHs or ALPs constitute a fraction of dark matter. 

In this \textit{letter}, motivated by the results from \cite{pavivek25}, we use the Lyman-$\alpha$ forest, a unique tracer of the matter density fluctuations in the underdense IGM, to search for a subdominant isocurvature contribution from ALPs produced after inflation to the primordial fluctuations that seed the growth of structure. We use the 1D Lyman-$\alpha$ flux power spectrum, which characterizes these matter perturbations, and is also affected at the small scales by complex gas dynamics arising from reionization. Using new simulations from the Sherwood-Relics project, we marginalize over these \textit{baryonic} physics and constrain the imprint of isocurvature perturbations on the flux power spectrum.   

\textit{Isocurvature modes in the matter power spectrum}---We consider models where the ALPs emerge at a symmetry breaking scale $f_{\rm{A}}$ after inflation, where $f_{\rm{A}}$ is fixed by the requirement that ALPs constitute all the dark matter energy density. After symmetry breaking, the axion field is essentially massless and offset from its vacuum value by a random angle $\theta_{\rm{ini}}$. When, $m_a$ $\approx$ $H$, the ALP starts oscillating at the temperature $T_{\rm{osc}}$ $<<$ $f_{\rm{A}}$ in the radiation era. The field then behaves like non-relativistic matter with energy density $\rho_a \propto  \theta_{\rm{ini}}$ \cite{irsic19, feix19}. 

Recall that during cosmic inflation the comoving event horizon decreases, leading to casually disconnected patches of the sky that were once connected. When the symmetry breaking takes place after inflation, regions larger than the horizon at $T=T_{\rm{osc}}$, defined as $k_{\rm{osc}} = aH_{{\rm{osc}}}$, are therefore uncorrelated.
Hence, the angle $\theta_{\rm{ini}}$ varies randomly across these regions. The Poissonian fluctuations in $\theta_{\rm{ini}}$ are then inherited by the axion number density leading to isocurvature perturbations in the matter power spectrum. In this way, the ALP field induces isocurvature perturbations in the dark matter fluid with dimensionless power spectrum, ${\Delta^2}_{\rm{iso}}(k) \propto k^{n_{\rm{iso}}-1}$, where $n_{\rm{iso}}$ is the isocurvature sprectral index. 
The dimensionless isocurvature power spectrum has subsequent evolution \cite{feix19},
\begin{equation} \label{ps}
\Delta_{\rm{iso}}^2 (k, z)= {D^2_{\rm{iso}}}(z)  {T^2_{\rm{iso}}}(k, z)A_{\rm{iso}}  \left( \frac{k}{k_{\star}} \right)^{n_{\rm{iso}}-1},
\end{equation}
 where $A_{\rm{iso}} = f_{\rm{iso}}^2 {A_{s}}$, $D_{\rm{iso}}$ is the growth factor accounting for the constant growth of isocurvature matter perturbations during the radiation era compared to the adiabatic case, and  $T_{\rm{iso}}$ is the isocurvature transfer function from \cite{bardeen1986}. Assuming uncorrelated isocurvature and adiabatic modes, the total power spectrum is consequently given by $\Delta^2_{\rm{total}} = \Delta_{\rm{ad}}^2 + \Delta_{\rm{iso}}^2$, where $\Delta_{\rm{ad}}^2$ is the usual dimensionless adiabatic power spectrum \cite{bardeen1986,peebles1970}.
 
More generally, a smaller (red-tilted) spectral index in Eq.~(\ref{ps}), increases the isocurvature contribution on large scales, which is a good approximation for axion models where the symmetry breaking occurs prior to inflation \cite{planck15}. The opposite applies to a larger (blue-tilted) $n_{\rm{iso}}$, arising in inflationary models proposed by \cite{kasuya19, chung18}. While the latest Planck  analysis does not show a preference for isocurvature modes with $n_{\rm{iso}}=2-4$ \cite{planck18_inflation}, the CMB is very weakly sensitive to this parameter given the $k$-range probed, the degeneracy with other cosmological parameters, and the fact that low-$l$ multipoles are extremely well fitted by adiabatic perturbations. The  ALPs in the post-inflation limit described above, however, provide a different mechanism for the $n_{\rm{iso}}$=4 case \cite{irsic19, feix19, fairbairn18}, which we keep fixed in this work. Another model that motivates the white-noise component are randomly distributed PBHs constituting dark matter and formed deep in the radiation era \cite{afshordi03, murgia19}. 

\begin{figure}[hbtp!]
    \centering
    \includegraphics[width=\linewidth]{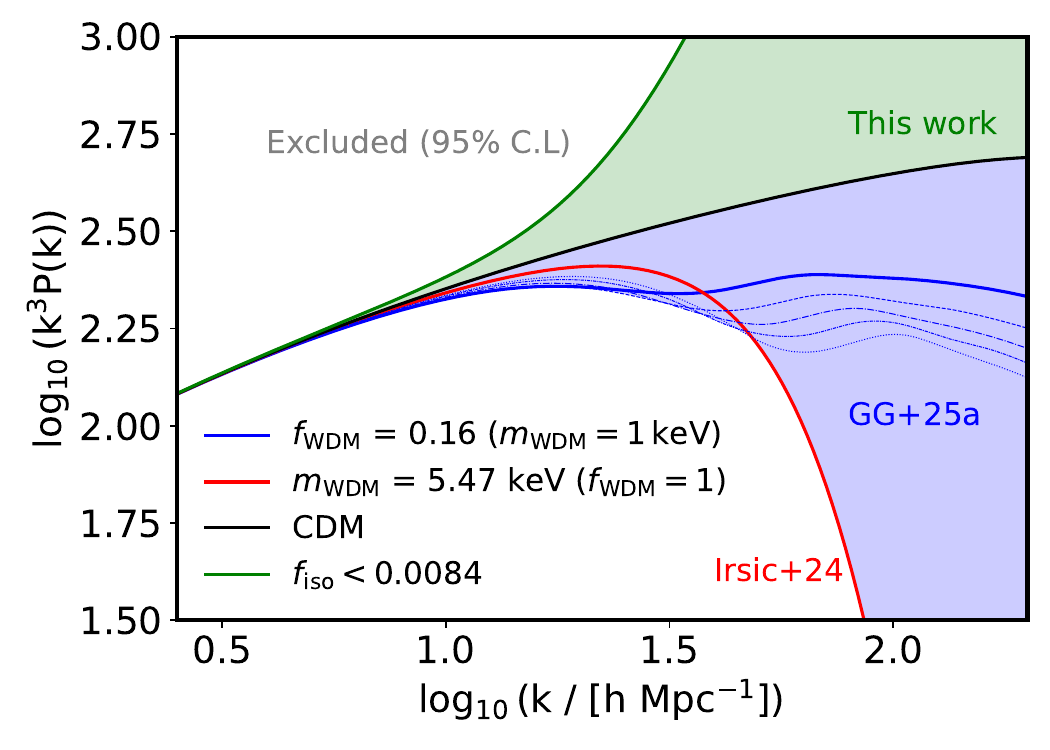}
    \caption{Constraints on the linear matter power spectrum at $z$=0. Bounds from below the CDM (black) line are inferred from WDM (\cite{irsic23}) and CWDM cosmologies (\cite{gg25}) in red and blue solid lines, respectively. For the latter, we further show in varying blue line styles the models that delineate the 2$\sigma$ allowed region, corresponding to the blue shaded area (see Fig.2 in \cite{gg25}).  Similarly, we show in green the bounds from this work above the CDM line. The solid green line corresponds to the fraction of isocurvature fluctuations allowed with a conservative noise treatment at 95\% C.L, and corresponding allowed region of models also in shaded green. }\label{matterPS}
\end{figure}

In Fig.~\ref{matterPS}, we show the isocurvature constraints on the linear matter power spectrum found in this work. We further show constraints from Cold+Warm dark matter (CWDM) and WDM cosmologies, where the free-streaming of the WDM component prevents structure formation below a scale that depends on the dark matter mass, suppressing the small-scale power relative to CDM \cite{irsic23, gg25}. 

\textit{Data}--- The 1D Lyman-$\alpha$ flux power spectra measurements are obtained from a sample of 15 high signal-to-noise ratio, high-redshift spectra from UVES and HIRES spectrographs presented in \cite{boera19}, with redshift coverage $z_{\rm{bin}}$=[4.2, 4.6, 5.0]. The \cite{boera19} data includes instrumental resolution correction and probes up to $k_{\rm{max}}$ = 0.2 s/km, extending  to a factor two  smaller scales than previous data could use in this context \cite{murgia19}. These small scales are the most subject to noise and metal contamination. The former is approximated as white noise and subtracted from the total flux power spectrum in \cite{boera19}, the latter typically only accounts for metals that leave an imprint redwards of the Lyman-$\alpha$ line (uncorrelated, e.g. CIV, SiIV). We model the contribution from SiIII cross-correlation following the prescription from \cite{ma25} and further assess noise mis-estimation similar to \cite{irsic23}. Capturing these effects allows us to robustly determine the power on $\sim$ 50-100 ckpc/h scales which can precisely pick up the isocurvature signature. 

\textit{Hydrodynamic Simulations}---The Sherwood-Relics suite models the non-linear physical processes of the IGM during reionization and allows to obtain synthetic Lyman-$\alpha$ forest spectra  to compare against the data \cite{puchwein23}. In previous works, these high-resolution simulations have allowed to constrain cosmologies beyond $\Lambda$CDM \cite{irsic23, gg25, pavivek25}. The set-up is very similar in this work, with a box size of 20$h^{-1}$Mpc and number of dark matter and gas particles $N_{\rm{p}}=1024^3$ for the reference simulation. Additional simulations with varying box size and particle number are used for the resolution and box size correction, defined as ${R_{s}}$ (see \cite{irsic23} for more details). Each reference run is complemented by 12 simulations where the reionization history is varied through the cumulative heat injected per proton $u_0(z)$ (see Table 1 in \cite{irsic23} for more details). We extend these models in post-processing to account for thermal broadening, parametrized by the temperature at mean density $T_0(z)$, and mean IGM transmission according to \cite{boera19}, given by $\tau_{\rm{eff}}(z)$. 

The initial conditions for the new isocurvature simulations are modified with the additional isocurvature component, following Eq. (\ref{ps}), according to which the particles are displaced at $z$=99. We run simulations for $f_{\rm{iso}}$=[0.001,0.005,0.01] at fixed $n_{\rm{iso}}$=4. For each $f_{\rm{iso}}$ we further simulate the 12 thermal history models discussed. The effect of the patchy reionization correction is included \textit{a posteriori} (see \cite{gg25, molaro22}). We then obtain the theoretical 1D flux power spectrum for each of these models and perform parameter inference using the \cite{boera19} data. The likelihood from \cite{boera19} is evaluated using a neural network emulator that follows the characteristics of that used in \cite{gg25}, reaching a 1\% level precision.  

\begin{figure}[hbtp!]
    \centering
    \includegraphics[width=0.95\linewidth]{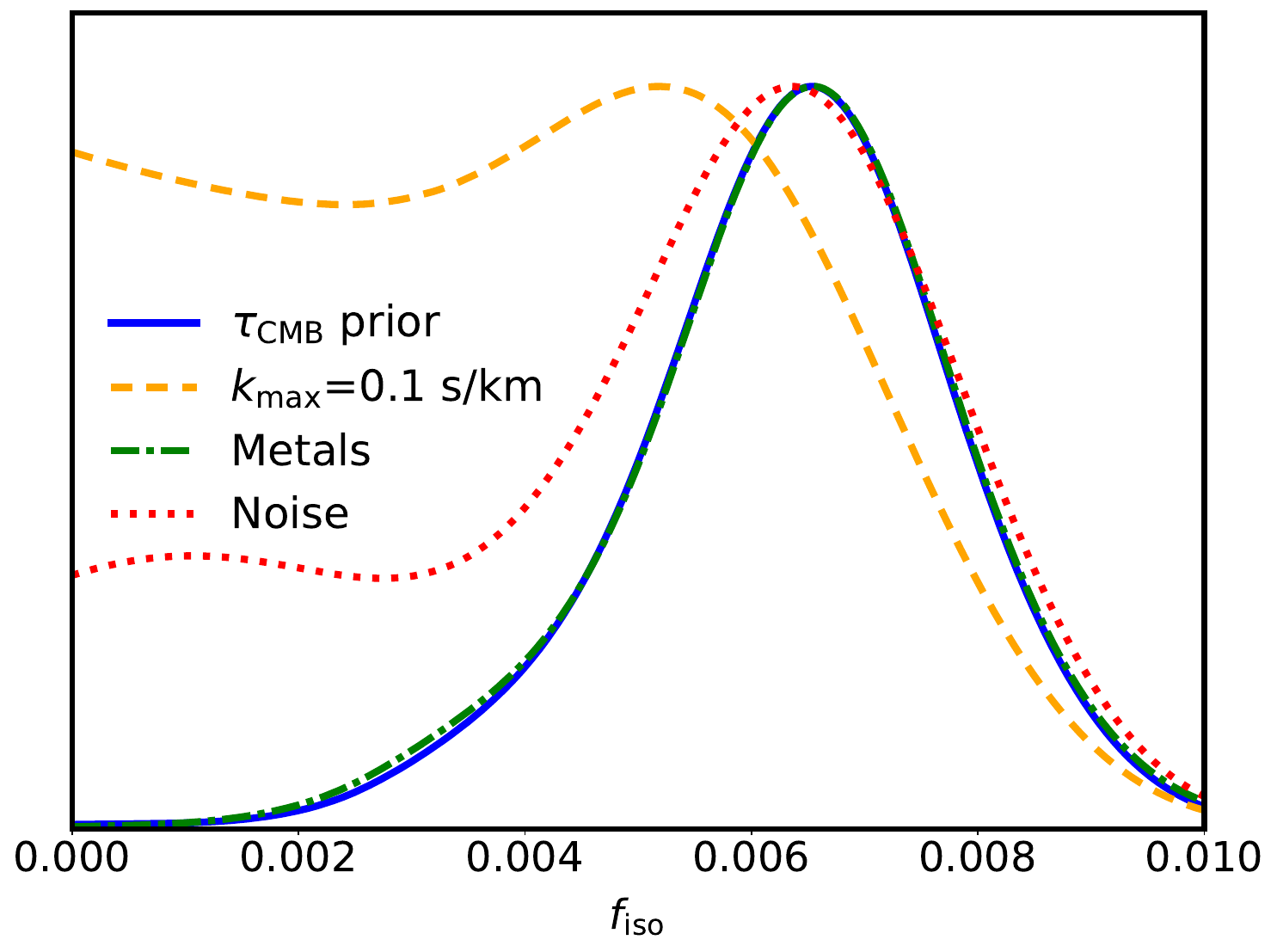}
    \caption{1D posterior distribution for $f_{\rm{iso}}$ for our default analysis ($\tau_{\rm{CMB}}$ priors) in blue. We also show the analysis where the last three $k-$bins are excluded in the dashed orange line. The dotted red line shows the analysis where a noise term in the flux power spectrum prediction per redshift bin is included, following \cite{irsic23}. For the analysis shown in dash-dot green, we include the contribution from Si-III, following \cite{ma25}.} \label{1d}
\end{figure}

\textit{Results.---} 
We parameterize the isocurvature fraction with $f_{\rm{iso}}$. The MCMC likelihood analyses use a conservative uniform prior: $f_{\rm{iso}} \in$ [0, 0.01], matching the range of isocurvature simulations run. We further use the mapping between the thermal parameters and the Thomson scattering optical depth $\tau_{\rm{e}}$, introduced by \cite{gg25b}, to inform the sampled combination (${T_0}^{z_i}$, ${u_0}^{z_i}$), by the latest $\tau_{\rm{e}}$ measurement from Planck, $\tau_{\rm{CMB}}$=0.054$\pm$0.007 \cite{planck18}. This  sets an effective prior in the $u_0-T_0$ envelope that is independent of cosmology, since the mapping with $\tau_{\rm{e}}$ is derived using phenomenological photo-ionization and heating models that match Lyman-$\alpha$ forest opacity observations \cite{puchwein23, keating20}. The different runs also include the numerical resolution correction, ${R_{s}}$, applied \textit{a posteriori} of the NN interpolation.

\begin{figure*}[hbtp!]
    \centering
    \includegraphics[width=0.78\linewidth]{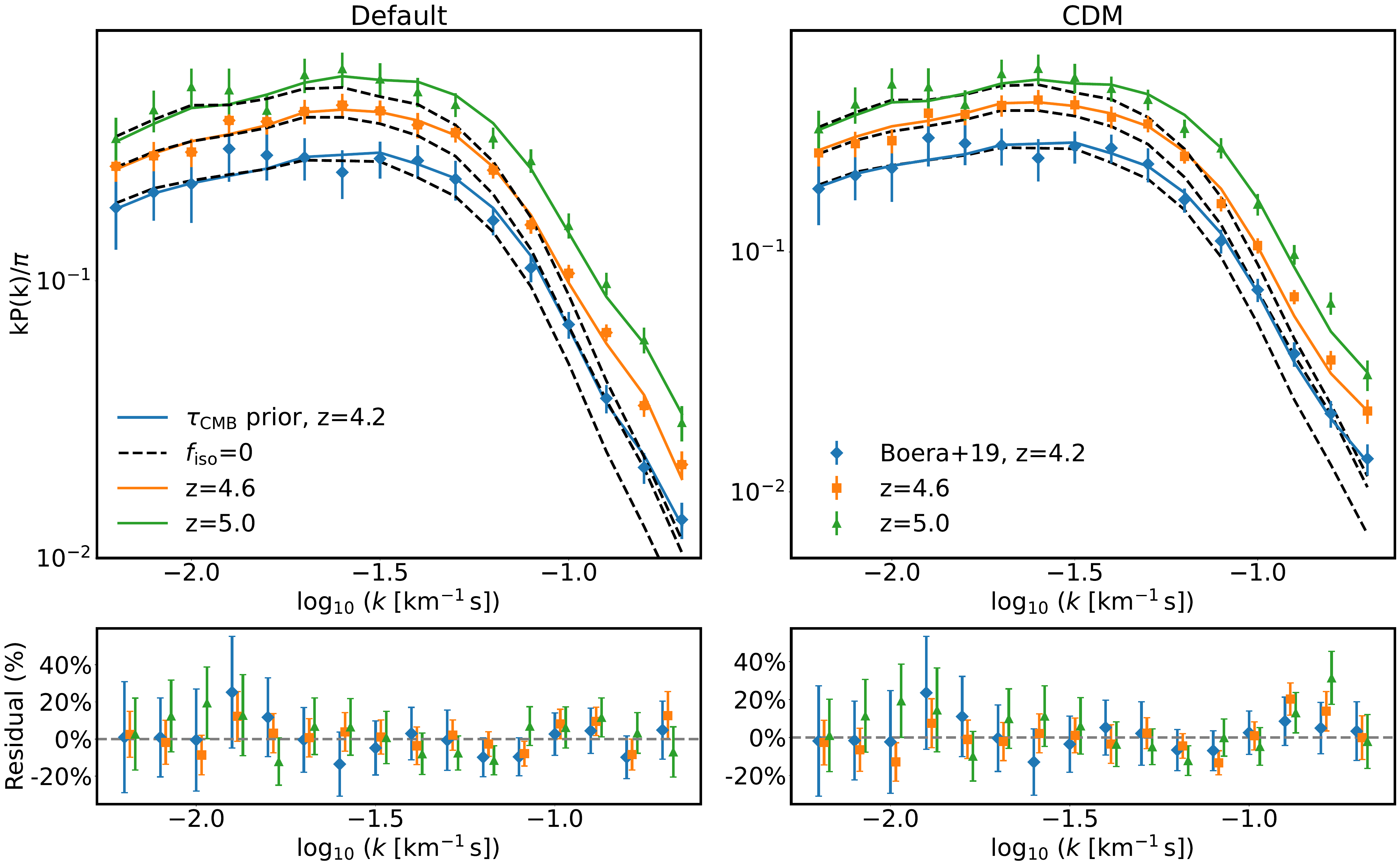}
    \caption{Best-fit plot at each redshift bin. The top \textit{left} panel shows the Default analysis in different colored lines. Both the top \textit{left} and \textit{right} panels show in dashed black lines the theory model with the same parameters as the Default analysis except for $f_{\rm{iso}}$=0. The top \textit{right} panel shows the best-fit from the CDM analysis with the same color coding for each redshift. The contours for this analysis in the $u_0^{5.0}-T_0^{5.0}$ and $u_0^{5.0}-\fiso$ planes are shown in Fig. \ref{u0t0} of the \textit{Supplemental Material}. The bottom panels shows the percentage residual of the data over each model for Default (\textit{left}) and CDM (\textit{right}). We have shifted the redshift points $z$=4.6 (orange) and $z$=5.0 (green) along the x-axis on the bottom two panels for visual clarity.}   \label{bestfit}
\end{figure*}

The main result is shown in Fig.~\ref{1d}, corresponding to the marginalized distribution of $f_{\rm{iso}}$ for the various analyses including Default, (with $\tau_{\rm{CMB}}$ prior + ${R_{s}}$), and consecutive analyses where we further set $k_{\rm{max}}$=0.1 s/km, and include the correction from  patchy reionization and Si-III, respectively. 
The default analysis yields $\fiso$ = ${0.0064^{+0.0012}_{-0.0014}}$ (68\% C.L) with $\chi^2$=21.7/35 excluding CDM by 4.6$\sigma$. In contrast, the $\chi^2$ increases for the CDM analysis ($\chi^2 = 31.9/36$), showing a similar level of improvement as in \cite{pavivek25} when fitting the \cite{boera19} data with a theory model that includes PMFs. 
The left-hand side plot in Fig.~\ref{bestfit} shows the best-fit model against the data at each redshift bin in colored lines. The black dashed lines in the top left and right panels show the fit with fixed astrophysical parameters (i.e. $T_0$($z$, $\gamma$($z$), $\tau_{\rm{eff}}$($z$) and $u_0$($z$)) and $f_{\rm{iso}}$=0, emphasizing that the isocurvature model provides a better fit than the CDM case, driven fundamentally by the last 3 $k$-bins. The fit from the CDM analysis and corresponding percentage residuals, shown also in Fig. \ref{bestfit} on the right column, shows limited agreement with the data towards these small scales, compared to the Default model. 

We check this further with a scale cut on the data such that $k_{\rm{max}}=0.1$ s/km, analysis shown by the dashed orange line in Fig.~\ref{1d}. The smallest scales are sensitive to the power enhancement from the constribution of isocurvature perturbations. Removing them, therefore, results on a broad $\fiso$ posterior with upper bound $\fiso< 0.0078$ (95\% C.L). 

These small scales are sensitive to noise modeling assumptions and metal contamination when computing the flux power spectrum measurements \cite{boera19}. Following \cite{irsic23}, we add a  noise term to the emulated flux power spectra that quantifies the offset of the noise flux power from that originally estimated by \cite{boera19}. This analysis, therefore, assumes a more conservative estimate of the noise, initially motivated both to asses noise misestimate in the measurements and to match the smallest scales in the \cite{boera19} data within CDM. We find that the noise term is degenerate with $\fiso$, weakening the upper bound to  $\fiso< 0.0084$ (95\% C.L), with the corresponding marginalized distribution shown by the dotted red line in Fig.~\ref{1d}. 

Furthermore, we account for the contribution of Si-III on small-scales using the prescription introduced by \cite{ma25}. The best-fit model (the dash-dot green line in Fig.~\ref{1d}) experiences no change compared to the Default analysis. We find that the contribution from SiIII correction at these scales is smaller than the resolution and box size correction which leads to very minor change in the fit, $\chi^2$=22/37, similar to the findings for WDM constraints \cite{ma25}.


When we include a thermal-dependent resolution and patchy correction, the $\fiso$ constraints remain largely unchanged, similar to \cite{pavivek25}'s findings ($\chi^2=23.3/35$ for the Patchy + $R_{s}(u_0)$ analysis shown in the \textit{Supplemental Material}).

We notice that a strong degeneracy between $\fiso-{u_0}(z)$ (see Fig.~\ref{u0t0} in the \textit{Supplemental Material}), which results in wide uncertainties on $u_0$, specially at the lower redshift bins of the data, (e.g.  ${u_0}^{4.6}=16.86^{+2.56}_{-5.48}$ eV/$m_{\rm{p}}$ in Default analysis). This is expected for models exhibiting more pressure smoothing, which can in turn be compensated with a higher isocurvature contribution.
Moreover, we also obtain slightly higher $T_0(z)$ values than \cite{boera19} ($T_{0}^{4.6}
= 0.79_{-0.13}^{+0.17}$ $10^4$K in the Default analysis), similar to the findings from \cite{irsic23} when external priors on $T_0$ parameter were imposed. The combined constraint on the IGM thermal history in our work aligns better with late-end point of reionization models strongly supported by Lyman-$\alpha$ opacity observations \cite{kulkarni19, keating20} than the constraints from \cite{boera19} for CDM cosmology.

In the framework of previous isocurvature constraints, these models have been fundamentally constrained with CMB anisotropies, with the most recent measurement $\fiso <$ 0.2 (95\% C.L) with fixed $n_{\rm{iso}}$=1 \cite{planck18_inflation}. Later constraints by \cite{feix19, feix20} for the $n_{\rm{iso}}$=4 case, indicate that the CMB prefers a non-zero $\fiso$, opposite to LSS probes (SZ+BAO). Their combined constraint, $\fiso <$0.64 (95\% C.L.), differs by two-orders of magnitude compared to our $\fiso$ constraints derived from the Lyman-$\alpha$ forest, mainly due to their limited sensitivity on the relevant scales.

Constraints on axion-like models are also commonly given in terms of $m_{\rm{a}}$, which depends on temperature as given in Eq.(2.2) in \cite{feix19}. The more familiar ALPs have a temperature-independent mass (i.e $n=0$). To translate our $\fiso$ limits to $m_{a}$ we solve the set of Eqs.(8)-(11) in \cite{irsic19}, fixing $A_{\rm{osc}}$ = 0.1 and $b$=1.  For the Noise analysis, this leads to $m_{\rm{a}} > 1.73 \times 10^{-18}$ eV (95\% C.L.). A slightly stronger bound was derived by \cite{irsic19} from PBHs constraints \cite{murgia19}. We constrain PBHs in reverse using the mapping from Eq.(5) in \cite{murgia19}. This allows us to derive the bound $f_{\rm{PBH}}M_{\rm{PBH}} < 650.6 M_\odot $ (95\% C.L.) from our $\fiso$ limit in the analysis with $k_{\rm{max}}$ = 0.1 s/km, which is the highest $k$-bin of the (previous) HIRES data used in that work. Our bound is notably lower (cf. $f_{\rm{PBH}}M_{\rm{PBH}} < 170 M_\odot)$ due partly to the poorer numerical resolution (with dark matter mass resolution $M_{\rm{DM}}=4.3\times10^{6}$\,h$^{-1}$$M_{\odot}$) of the simulations used in that work, and mainly to their IGM thermal evolution parameterization, where they assume a power law for $T_0(z)$. Our treatment of the thermal evolution is more conservative where we span a wide range of IGM thermal histories (see Table 1 in \cite{irsic23}). 

At the smallest end of the mass scale of ALPs, the candidate of interest is \textit{fuzzy dark matter}, whose effect on structure growth is instead to suppress the small-scale power due to quantum pressure \cite{marsh15}. While probing different physics, the Lyman-$\alpha$ forest allows $m_{a} >$ $10^{-20}$ eV \cite{irsic17, rogers21}, complementary to our results given the large mass range spanned  by ALPs.

The features for the axion models discussed above are built on the description of the more familiar QCD axion, whose $m_{a} $ has a temperature dependence set by non-perturbative effects that mimics the $n=4$ case \cite{feix19}.  Our $\fiso$ bound translates into $m_{\rm{a}}> 8.30\times 10^{-16} $eV, corresponding to an ultra-light QCD-like axion with correspondingly non-trivial decay constant $f_{\rm{A}}$ beyond the Planck scale and very small coupling to the Standard Model \cite{banks03}, which would in turn be disallowed by quantum gravity \cite{nima07}. Investigating the theoretical insights for such axion models is beyond the scope of this work, instead we mention constraints from different theoretical frameworks as separate phenomenological models that describe a generic axion as a dark matter candidate.




Other probes of the matter distribution at mildly non-linear scales, such as the Ultraviolet Luminosity Function (ULVF) of galaxies, have constrained ALPs \cite{gorghetto25, irsic19}, especially of interest now with new UV bright galaxy candidates at $z\approx25$ in JWST data \cite{perez25}. The most recent work in this context  found slightly stronger $m_{a}$ constraints (cf. $m_{\rm{a}} > 6.6 \times 10^{-17}$ eV \cite{urrutia25}). However, these constraints are  sensitive to the underlying galaxy-formation astrophysics, such as feedback and start-formation prescriptions, as well as the assumed selection function. 
In comparison, since the Lyman-$\alpha$ forest lies at the intersection where small-scales have not grown non-linearly yet, the \textit{astrophysics} to model, namely the cooling/heating of the gas, are comparatively straightforward.



\textit{Conclusions.---} We have run new simulations based on the Sherwood-Relics suite which include the effect of post-inflationary ALPs to fit the \cite{boera19} data. These ALPs induce isocurvature perturbations in the power spectrum, which enhance the power on small-scales, similar to PMFs \cite{pavivek25}. We parameterize the isocurvature component with $\fiso$, the ratio of isocurvature to adiabatic peturbations. We marginalize over  IGM thermal parameters within a Bayesian inference analysis framework with a newly trained emulator for this cosmology (c.f \cite{gg25}). We find a tentative detection with $\fiso$ = $0.0064^{+0.0012}_{-0.0014}$ (68\% C.L) for our baseline analysis with the published noise of the \lya\ forest data. Assessing the significance of this  results suggests scrutinizing the published estimate of the noise that affects the small scales of the data. We find that a more conservative estimate  of this noise weakens the upper bound to $\fiso< 0.0084$ (95\% C.L). Interestingly, the isocurvature model provides a better fit to the \cite{boera19} data than the CDM case in both analyses, similar to \cite{pavivek25}'s results for models with PMFs.
The bounds derived from the CMB are not competitive with our constraints, while the limits from the ULVFs of high-redshift galaxies are an order of magnitude stronger, but are dependent on complex astrophysics modelling. We have also updated previous Lyman-$\alpha$ forest  constraints on ALPs which were  translated from PBHs  bounds with lower ${k_{\rm{max}}}$ data . The $\sim$2 factor difference in $\fiso$ limits arises primarily from the assumed thermal history in that work and the numerical resolution. The impact of isocurvature cosmology on the thermal evolution of the IGM has not been studied, which motivates us to use a prior (on $\taureio$) agnostic to this cosmology, resulting in more conservative $\fiso$ limits.  

If correct, our tentative detection would leave a measurable imprint on the underlying dark matter distribution that could be  confirmed by upcoming CMB experiments (CMB-S4 \cite{cmbs4}), galaxy surveys (Euclid \cite{euclid}) and HI intensity mapping (SKA \cite{ska}). It would also leave an  imprint on the high-redshift UVLF that can be constrained with future JWST observations \cite{urrutia25}. The results from this work further highlight the potential of forthcoming small-scale Lyman-$\alpha$ forest data. Stronger bounds on axion dark matter will become possible  with the larger high-redshift quasar sample size and improved systematics achieved with the forthcoming GHOSTLy survey \cite{artola24, kalari24}. The complementarity of all the probes mentioned, each targeting a different redshift and scale regime, motivates  a multi-probe analysis that could potentially disentangle the signal from these ALPs, providing fundamental insights into the physics of the Early Universe.

\textit{Acknowledgments.---}%
The authors thank Takeshi Kobayashi and Jussi Valiviita for useful discussions. 
VI acknowledges partial support by the Kavli Foundation. The simulations used in this work were performed using the Cambridge Service for Data Driven Discovery (CSD3), part of which is operated by the University of Cambridge Research Computing on behalf of the STFC DiRAC HPC Facility (www.dirac.ac.uk).  The DiRAC component of CSD3 was funded by BEIS capital funding via STFC capital grants ST/P002307/1 and ST/R002452/1 and STFC operations grant ST/R00689X/1.  DiRAC is part of the National e-Infrastructure. Support by ERC Advanced Grant 320596 ‘The Emergence of Structure During the Epoch of Reionization’ is gratefully acknowledged. MGH has been supported by STFC consolidated grants ST/N000927/1 and ST/S000623/1. JSB is supported by STFC consolidated grant ST/X000982/1. MV is supported by IDEAS SISSA grant and INFN INDARK.


\bibliography{references}

\onecolumngrid
\appendix

\clearpage

\setcounter{equation}{0}
\setcounter{figure}{0}
\setcounter{table}{0}
\setcounter{page}{1}
\makeatletter
\counterwithin{figure}{section}
\renewcommand{\theequation}{S\arabic{equation}}
\renewcommand{\thefigure}{S\arabic{figure}}
\renewcommand{\thepage}{S\arabic{page}}
\renewcommand{\bibnumfmt}[1]{[S#1]}
\renewcommand{\citenumfont}[1]{S#1}

\begin{center}

\textbf{\large Post-inflationary axion constraints from the Lyman-$\alpha$ forest} 

\vspace{0.05in}

\textit{\large Supplemental Material} \

{Olga Garcia-Gallego, Vid Ir\v{s}i\v{c}, Matteo Viel, Martin Haehnelt \& James S. Bolton}

\end{center}

\begin{figure}[hbtp!]
    \centering
    \includegraphics[width=0.8\linewidth]{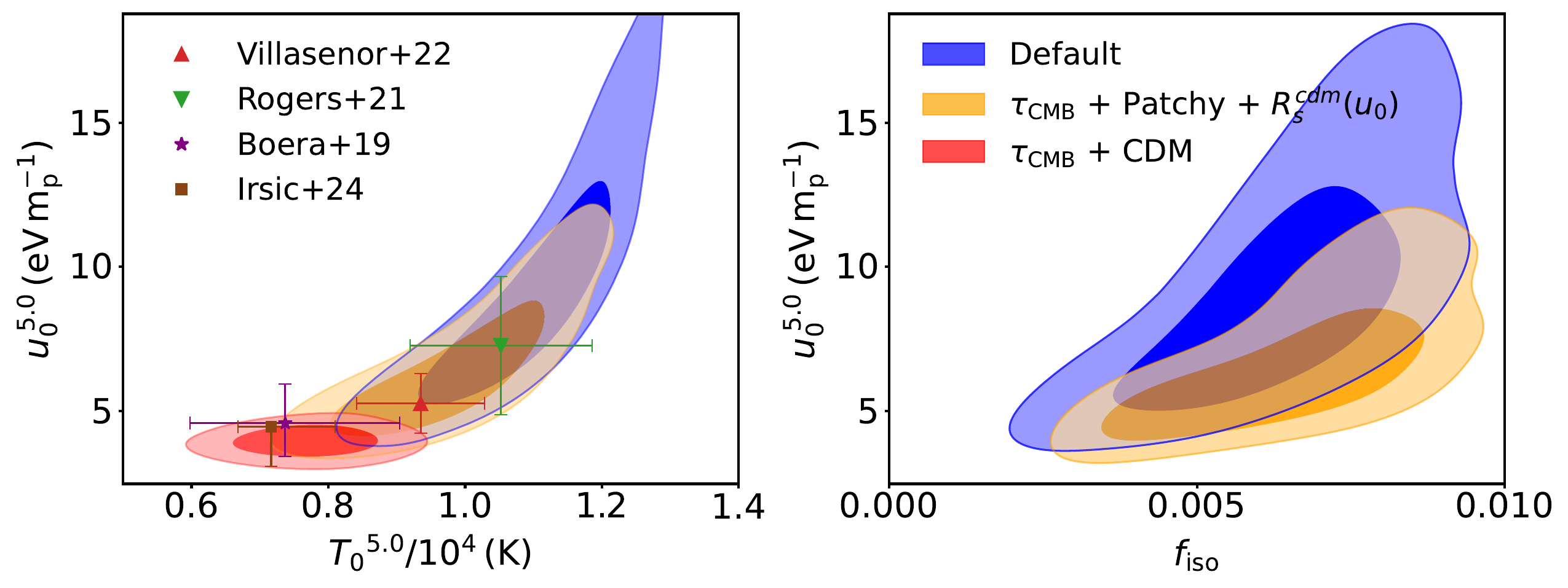}
    \caption{2D posterior in ${u_0}-{T_0}$ (\textit{left}) and ${u_0}-{\fiso}$ plane (\textit{right}) at $z$= 5.0, for the Default analysis (with $\tau_{\rm{CMB}}$ priors + $R_{s}$) in blue, with patchy and thermal-dependent resolution correction in orange, and for $\fiso =0$ (CDM) in red. Constraints for the same parameters from \cite{villasenor22} in red, \cite{rogers21} in green, \cite{boera19} in purple and \cite{irsic23} in brown are also included.} \label{u0t0}
\end{figure}

Figure \ref{u0t0} shows the 2D contours in the $u_0-T_0$ and $u_0-\fiso$ plane at $z$=5.0 for several analyses mentioned in the main text. The blue contours correspond to the default analysis (with $\tau_{\rm{CMB}}$ prior and resolution correction). The orange contour includes the effect of patchy reionization \cite{molaro22}, as well as of a thermal-dependent resolution correction \cite{irsic23}. The constraint                              obtained from the marginal $\fiso$ posterior in this case is $\fiso$=${0.0066^{+0.0015}_{-0.0018}}$ (68\% C.L.). The patchy and resolution correction have opposite effects of decreasing and increasing the small scale power respectively, which is why the $\fiso$ median remains largely unchanged, while the posterior broadens since the thermal parameters' contours become tighter (as discussed by \cite{irsic23}). We further show in red the contours obtained for $\fiso$=0 analysis. This is equivalent to the $\tau_{\rm{CMB}}$ prior labeled analysis in Figure S1 in \cite{gg25b}. 
The shape of the $u_0-T_0$ at ($z$=5.0) is driven by the $\tau_{\rm{CMB}}$ prior.
This can be understood by looking at the isocontours for $\taureio$=0.06 in Figure 2 of \cite{gg25b}. One can see that the $u_0$ parameter remains largely unconstrained (cf. the CDM analysis) given its large degeneracy with $\fiso$ shown in Figure \ref{u0t0} on the right-hand side. The same applies to $u_0$ parameter at lower redshifts $z$=4.2, 4.6. At these redshifts, the $u_0-T_0$ contours resemble those found in the context of warm dark matter constraints by \cite{irsic23} when no prior on the IGM thermal history is imposed.

\end{document}